Superconductivity in layered Oxychalcogenide $La_2O_2Bi_3AgS_6$

Rajveer Jha[1,*], Yosuke Goto[1], Ryuji Higashinaka[1], Tatsuma D. Matsuda[1], Yuji Aoki[1], and Yoshikazu Mizuguchi [1, †]

[1]Department of Physics, Tokyo Metropolitan University, 1-1 Minami-osawa, Hachioji, Tokyo 192-0397, Japan

## Abstract:

We report the superconductivity in layered oxychalcogenide $La_2O_2Bi_3AgS_6$ compound. The $La_2O_2Bi_3AgS_6$ compound has been reported recently by our group, which has a tetragonal structure with the space group *P*4/*nmm*. The crystal structure of $La_2O_2Bi_3AgS_6$ can be regarded as alternate stacks of $LaOBiS_2$-type layer and rock-salt-type (Bi,Ag)S layer. We measured low-temperature electrical resistivity and observed superconductivity at 0.5 K. The observation of superconductivity in the $La_2O_2Bi_3AgS_6$ should provide us with the successful strategy for developing new superconducting phases by the insertion of a rock-salt-type chalcogenide layer into the van der Waals gap of $BiS_2$-based layered compound like $LaOBiS_2$.

E-mail for corresponding authors: *rajveerjha@gmail.com, † mizugu@tmu.ac.jp

The discovery of $BiS_2$-based superconductors made a huge interest in the scientific community [1,2]. In this progress several new materials possessing the bismuth chalcogenides [$BiCh_2$ (Ch=S, Se)] layers have been designed [1-10]. Firstly, the superconductivity has been reported in $Bi_4O_4S_3$ layered compound with a transition temperature ($T_c$) of 4.5 K [1,3]. Later, other $BiS_2$-based superconductors $REO_{1-x}F_xBiS_2$ (RE = La, Ce, Nd, Yb, Pr) and $Sr_{1-x}La_xFBiS_2$ were discovered after $Bi_4O_4S_3$ [1–11]. So far, the superconductivity in the $BiS_2$ based compounds has been reported to be induced via charge carrier doping into $BiS_2$ layers for the $REO_{1-x}F_xBiS_2$ and $Sr_{1-x}RE_xFBiS_2$ compounds [3-11]. These superconductors are very sensitive to the external hydrostatic pressure and the highest $T_c$ above 10 K was achieved for $LaO_{0.5}F_{0.5}BiS_2$ and $Sr_{1-x}RE_xFBiS_2$ (RE = La, Ce,Nd, Pr, Sm) compounds [12-16].

Recently, new layered oxysulfide $LaOPbBiS_3$ has been reported by Y.-L Sun *et. al* [17]. They proposed that $LaOPbBiS_3$ compound is a narrow gap semiconductor with an activation energy of ∼17 meV, which suggested that $LaOPbBiS_3$ and its derivatives may be promising for thermoelectric applications [17]. After careful structural analysis for the $LaOBiPbS_3$ phase [18], we have proposed a material design strategy focusing on the van der Waals gaps between the two $BiCh_2$ layers. The van der Waals gaps of $LaOBiS_2$ can be filled by inserting some rock salt layers composed of $M_2S_2$ [M = Pb, Bi] [18], which results in the formation of $La_2O_2M_4S_6$-type compounds. Later on, Y. Hijikata et al. reported the synthesis of $La_2O_2Bi_3AgS_6$ compound [19]. The structure of $La_2O_2Bi_3AgS_6$ is similar to $LaOBiPbS_3$, but the interlayer bond (M2–S1 bond) in $La_2O_2Bi_3AgS_6$ is shorter than that of $LaOBiPbS_3$. The electrical conductivity and carrier concentration of $La_2O_2Bi_3AgS_6$ are higher than those of $LaOBiPbS_3$ and $LaOBiS_2$, possibly due to the shorter interlayer bonds and/or chemical pressure effect with compressed *a*-axis [19]. Our previous electrical resistivity and magnetization measurements showed no signature of a superconducting transition down to 2 K [19].

Here, we measured the electrical resistivity of $La_2O_2Bi_3AgS_6$ down to 0.1 K using an adiabatic demagnetization refrigerator (ADR) system. We observed superconductivity in $La_2O_2Bi_3AgS_6$ compound at $T_c^{zero}$ = 0.5 K. This is the first observation of superconductivity among $La_2O_2M_4S_6$-type compounds.

The polycrystalline sample of $La_2O_2Bi_3AgS_6$ was prepared by a solid-state reaction method. Powders of $Bi_2O_3$ (99.9%), $La_2S_3$ (99.9%), and AgO (99.9%) and grains of Bi (99.999%) and S (99.99%) with a nominal composition of $La_2O_2Bi_3AgS_6$ were mixed in a pestle and mortar, pelletized, sealed in an evacuated quartz tube, and heated at 720 °C for 15 h. The obtained sample was reground for homogeneity, pelletized, and heated at 720 °C for 15 h. The phase purity of the prepared sample and the optimal annealing conditions were examined using laboratory X-ray diffraction (XRD) with Cu-$K_\alpha$ radiation. The crystal structure parameters were refined using the Rietveld method with RIETAN-FP [20]. Schematic image of the crystal structure were drawn using VESTA (see the inset of Fig. 1) [21]. The electrical resistivity down to $T$ = 0.1 K was measured by four probe technique using adiabatic demagnetization refrigerator (ADR) system on the Physical Property measurement system (PPMS: Quantum Design).

Figure 1 shows the room temperature XRD pattern of $La_2O_2Bi_3AgS_6$ compound. The $La_2O_2Bi_3AgS_6$ compound crystallized in the tetragonal structure with the space group of *P*4/*nmm*, which is composed of stacked [$M_4S_6$] layers and fluorite-type [$La_2O_2$] layers. No impurity peak has been observed, and all the observed peaks are well indexed to the main phase of $La_2O_2Bi_3AgS_6$ compound, which is comparable to the previous study [19]. The estimated lattice parameters are $a$ = 4.0568(1) Å and $c$ = 19.348(1) Å. The crystal structure of $La_2O_2Bi_3AgS_6$ compound is shown in the inset of Fig. 1, in which chalcogenide layers of (BiAg)S and LaO(BiAg)$S_2$ are alternately stacking.

Figure 2 shows the temperature dependence of electrical resistivity [$\rho(T)$] from 300 to 0.1 K. The overall behavior of the normal state electrical resistivity is metallic. The lower resistivity in comparison with previous report might be due to the smaller lattice parameters for this sample. The smaller lattice parameter $c$ can make the shorter interlayer M1-S3 bonding. Due to the smaller lattice parameter $a$ the in-plane M1-S1 bonding becomes shorter, which is preferable for metallic conductivity [22]. We have analyzed the composition of the sample using energy dispersive x-ray spectroscopy (EDX) and found slightly high concentration of Ag in the present sample in comparison with the previous sample [19]. The smaller lattice parameters might be due to the excess Ag concentration in the present sample.

Notably, we observed a broad hump in the $\rho(T)$ curve below $T^*\sim$180 K, which is quite similar to the hump observed in the $\rho(T)$ curve below ~280 K for the $EuFBiS_2$ compound, which shows superconductivity at 0.3 K [23]. The origin of the hump in the $\rho(T)$ curve was proposed as the charge density wave (CDW) transition in $EuFBiS_2$. From the analogy, we assume that the anomaly is caused by a CDW transition.

The resistivity starts to deviate from the normal state value below the temperature 2.5 K and becomes zero at 0.5 K [see the inset of Fig. 2], which suggests the emergence of superconductivity in the $La_2O_2Bi_3AgS_6$ compound. As mentioned above, the anomaly at $T^*$ may be related to a CDW transition. Therefore, it is expected that some part of the Fermi surface disappears in the CDW state and the superconductivity develops on the remaining part. With the $La_2O_2Bi_3AgS_6$ and related $La_2O_2M_4S_6$ systems, we may be able to discuss the universal relationship between superconductivity and CDW in the $BiCh_2$-based compounds.

Furthermore, there is another notable characteristics in the $La_2O_2M_4S_6$ system. According to the band calculations for $LaOBiPbS_3$ [18], the valence bands just below the Fermi energy ($E_F$) are mainly composed of S-3p orbitals of the S3 site (see the inset of Fig. 1) in the $M_2S_2$ layer. This is different from the case of the typical $BiS_2$-based compounds in which the valence bands just below $E_F$ are composed of S-3p orbitals of the in-plane S1 site. The similar situation is expected for $La_2O_2Bi_3AgS_6$ compound: the valence bands near $E_F$ would be composed of S-3p bands of the rock-salt type (Bi,Ag)S layer. Therefore, the superconductivity states in $La_2O_2Bi_3AgS_6$ may be different from those of the typical $BiS_2$-based superconductors because of the difference in the electronic states near $E_F$. We expect that the discovery of the superconductivity in $La_2O_2Bi_3AgS_6$ will extend the field of layered bismuth chalcogenide superconductors.

In conclusion, we reported the observation of superconductivity ($T_c^{zero}$ = 0.5 K) in the polycrystalline sample of layered oxychalcogenide $La_2O_2Bi_3AgS_6$. In the temperature dependence of resistivity, an anomaly possibly related to the CDW transition ($T^* \sim 180$ K) is observed. The $La_2O_2Bi_3AgS_6$ and related $La_2O_2M_4S_6$ systems will be useful to discuss the universal relationship between superconductivity and CDW states and to develop new CDW-related superconductors.

Acknowledgments:

We gratefully appreciate K. Matsubayashi and Y. Yuan of University of Electrocommunication and O. Miura and Y. Hijikata of Tokyo Metropolitan University for fruitful discussions. This work was financially supported by grants in Aid for Scientific Research (KAKENHI) (Grant Nos. 15H05886, 15H05884, 16H04493, 17K19058, 16K05454, and 15H03693)

Reference:

1. Y. Mizuguchi, H. Fujihisa, Y. Gotoh, K. Suzuki, H. Usui, K. Kuroki, S. Demura, Y. Takano, H. Izawa, and O. Miura, Phys. Rev. B **86**, 220510 (2012).
2. Y. Mizuguchi, S. Demura, K. Deguchi, Y. Takano, H. Fujihisa, Y. Gotoh, H. Izawa, and O. Miura, J. Phys. Soc. Jpn. **81**, 114725 (2012).
3. S. K. Singh, A. Kumar, B. Gahtori, S. Kirtan, G. Sharma, S. Patnaik, and V. P. S. Awana, J. Am. Chem. Soc. **134**, 16504 (2012).
4. S. Demura, Y. Mizuguchi, K. Deguchi, H. Okazaki, H. Hara, T. Watanabe, S. J. Denholme, M. Fujioka, T. Ozaki, H. Fujihisa, Y. Gotoh, O. Miura, T. Yamaguchi, H. Takeya,Y. Takano, J. Phys. Soc. Jpn. **82**, 033708 (2013).
5. R. Jha, A. Kumar, S. K. Singh, V. P. S. Awana, J. Supercond. Nov. Magn. **26**, 499 (2013)
6. J. Xing, S. Li, X. Ding, H. Yang, H.H. Wen, Phys. Rev. B **86**, 214518 (2012).
7. D. Yazici, K. Huang, B. D. White, A. H Chang, A. J. Friedman, M. B. Maple, Philos. Mag. **93**, 673 (2012).
8. D. Yazici, K. Huang, B. D. White, I. Jeon, V. W. Burnett, A.J. Friedman, I. K. Lum, M. Nallaiyan, S. Spagna, M. B. Maple, Phys. Rev. B **87**, 174512 (2013).
9. A. Krzton-Maziopa, Z. Guguchia, E. Pomjakushina, V. Pomjakushin, R. Khasanov, H. Luetkens, P. Biswas, A. Amato, H. Keller, K. Conder, J. Phys.: Condens. Matter **26**, 215702 (2014).
10. Y. Mizuguchi, A. Omachi, Y. Goto, Y. Kamihara, M. Matoba, T. Hiroi, J. Kajitani, O. Miura, J. Appl. Phys. **116**, 163915 (2014).
11. X. Lin, X. Ni, B. Chen, X. Xu, X. Yang, J. Dai, Y. Li, X. Yang, Y. Luo, Q. Tao, G. Cao, Z. Xu, Phys. Rev. B **87**, 020504 (2013).
12. H. Kotegawa, Y. Tomita, H. Tou, H. Izawa, Y. Mizuguchi, O. Miura, S. Demura, K. Deguchi, Y. Takano, J. Phys. Soc. Jpn. **81**, 103702 (2012).
13. C. T. Wolowiec, D. Yazici, B. D. White, K. Huang, M. B. Maple, Phys. Rev. B **88**, 064503 (2013).
14. C. T. Wolowiec, B. D. White, I. Jeon, D. Yazici, K. Huang, M. B. Maple, J. Phys.: Condens. Matter **25**, 422201 (2013).
15. R. Jha, B. Tiwari, V. P. S. Awana, J. Phys. Soc. Jpn. **83**, 063707 (2014).

16. R. Jha, B. Tiwari, V. P. S. Awana, J. Appl. Phys. **117**, 013901 (2015).
17. Y.-L. Sun, A. Ablimit, H.-F. Zhai, J.-K. Bao, Z.-T. Tang, X.-B. Wang, N.-L. Wang, C.-M. Feng, and G.-H. Cao, Inorg. Chem. **53**, 11125 (2014).
18. Y. Mizuguchi, Y. Hijikata, T. Abe, C. Moriyoshi, Y. Kuroiwa, Y. Goto, A. Miura, S. Lee, S. Torii, T. Kamiyama, C. H. Lee, M. Ochi, K. Kuroki, EPL **119** 26002 (2017).
19. Y. Hijikata, T. Abe, C. Moriyoshi, Y. Kuroiwa, Y. Goto, A. Miura, K. Tadanaga, Y. Wang, O. Miura, and Y. Mizuguchi, J. Phys. Soc. Jpn. **86**, 124802 (2017).
20. F. Izumi and K. Momma, Solid State Phenom. **130,** 15 (2007).
21. K. Momma and F. Izumi, J. Appl. Crystallogr. **41**, 653 (2008).
22. Y. Mizuguchi, A. Miura, J. Kajitani, T. Hiroi, O. Miura, K. Tadanaga, N. Kumada, E. Magome, C. Moriyoshi, Y. Kuroiwa, Sci. Rep. **5,** 14968 (2015).
**23.** H.-F. Zhai, Z.-T. Tang, H. Jiang, K. Xu, K. Zhang, P. Zhang, J.-K. Bao, Y.-L. Sun, W.-H. Jiao, I. Nowik, I. Felner, Y.-K.Li, X.-F. Xu, Q. Tao, C.-M. Feng, Z.-A. Xu, and G.-H. Cao, Phys. Rev. B **90**, 064518 (2014).

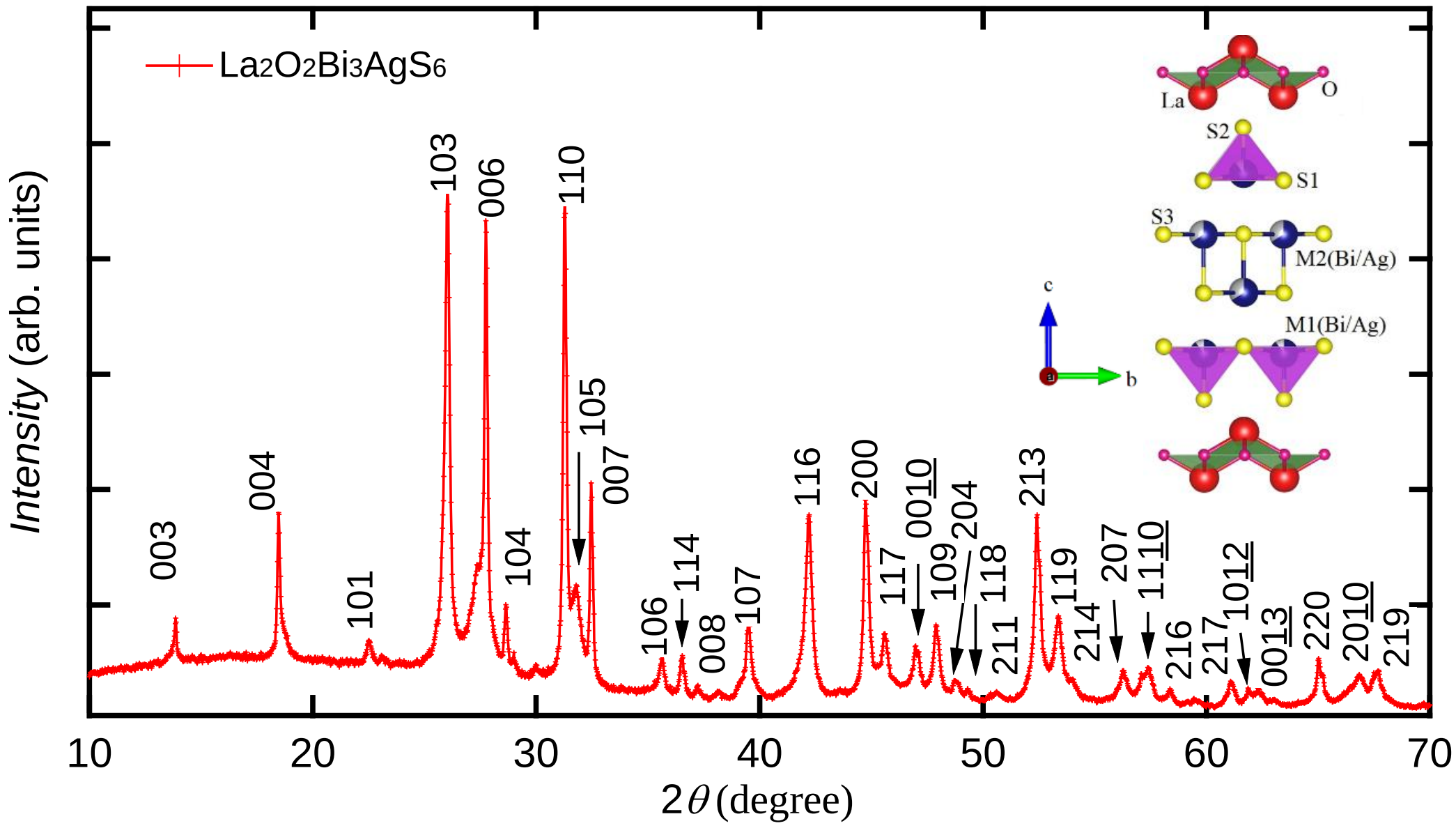

**Fig. 1:** (color online) The room temperature XRD pattern of $La_2O_2Bi_3AgS_6$ compound. The numbers in the XRD pattern are Miller indices. Inset is the schematic unit cell obtained by a refinement using the Rietveld method with RIETAN-FP [20]

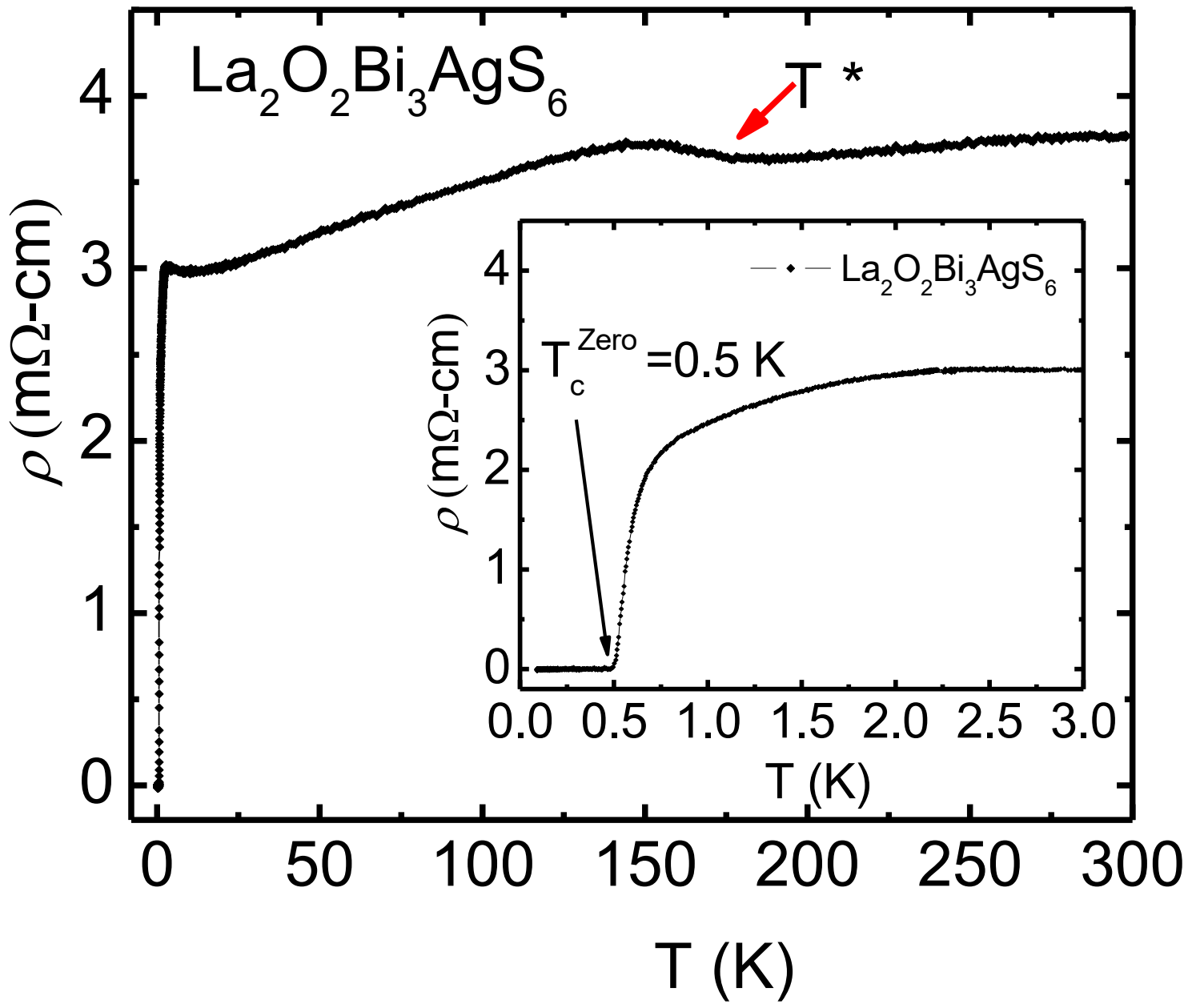

**Fig. 2:** (color online) The temperature dependence of electrical resistivity down to 0.1 K for the $La_2O_2Bi_3AgS_6$ compound. A hump structure appearing in the $\rho(T)$ curve is denoted by $T^*$. Inset is the $\rho(T)$ curve in the temperature range 3.0 to 0.1 K.